\documentclass{jfm}

\usepackage[applemac]{inputenc}
\usepackage[usenames,dvipsnames,svgnames,table]{xcolor}
\usepackage{graphicx}

\usepackage{natbib}%

\usepackage{amsbsy}
\usepackage{amssymb}
\usepackage{amsmath}

\usepackage{psfrag}

\newcommand{\reff}[1]{\ref{fig:#1}} 
\newcommand{\refs}[1]{\ref{sec:#1}} 
\newcommand{\refss}[1]{\ref{subsec:#1}} 

\newcommand{\ie}{\textit{i.e.}}
\newcommand{\del}{{\Delta^*}}

\usepackage{pdfcomment}

\title[Droplet impact near a millimetre-sized hole]{Exploring droplet impact near a millimetre-sized hole: comparing a closed pit with an open-ended pore}

\author[R. de Jong, O.R. Enr\'iquez, and D. van der Meer]{R\ls I\ls A\ls N\ls N\ls E\ns D\ls E\ns J\ls O\ls N\ls G\footnote{Email address for correspondence: jong.riannede@gmail.com},\ns O\ls S\ls C\ls A\ls R\ns R.\ns E\ls N\ls R\ls I\ls Q\ls U\ls E\ls Z,\ns \break \and D\ls E\ls V\ls A\ls R\ls A\ls J\ns V\ls A\ls N\ns D\ls E\ls R\ns M\ls E\ls E\ls R}
\affiliation{Physics of Fluids Group, MESA+ Institute for Nanotechnology, and J.M. Burgers Centre for Fluid Dynamics, University of Twente, P.O. Box 217, 7500 AE Enschede, The Netherlands}
\date{\today}

\begin{document}
\maketitle


\begin{abstract}
We investigate drop impact dynamics near closed pits and open-ended pores experimentally. The resulting impact phenomena differ greatly in each case. For a pit, we observe three distinct phenomena, which we denote as: a splash, a jet and an air bubble, whose appearance depends on the distance between impact location and pit. Furthermore, we found that splash velocities can reach up to seven times the impact velocity. Drop impact near a pore, however, results solely in splashing. Interestingly, two distinct and disconnected splashing regimes occur, with a region of planar spreading in-between. For pores, splashes are less pronounced than in the pit case. We state that, for the pit case, the presence of air inside it plays the crucial role of promoting splashing and allowing for air bubbles to appear.
\end{abstract}

\section{Introduction}

Droplet impact can be observed regularly in daily life, for example, when raindrops hit the ground or while washing the dishes. Additionally, in industrial processes such as spray coating, spray cooling and ink-jet printing, drop impact is of major importance. The topic has been studied for more than a century, starting with Worthington, who in 1876 listed the huge variety in shapes that the drop can take after impact on a solid substrate \citep{Worthington1876}. It has been found that the liquid properties, the droplet size and velocity, the wettability and roughness of the substrate, as well as the surrounding gas pressure, all have an influence on the behaviour of the drop after impact. For example, under different conditions the lamella spreading may lead to prompt splashing, corona splashing and fingering \citep{Rioboo2001, Yarin2006, Richard2002, Clanet2004, Xu2005, Tsai2009, Bouwhuis2012}.

In spite of the fact that many real surfaces are likely to be rough and inhomogeneous, droplet impact research has mostly concentrated on homogeneous isotropic solid surfaces. However, some authors have looked at smooth solid substrates containing a single obstacle \citep{Delbos2010, Ding2012a, Lorenceau2003, Roisman2010, Josserand2005}, \emph{e.g.}, the impact of a droplet near an open-ended hole (pore), revealing different flow behaviour depending on the impact location. For a centred impact, the inner part of the droplet will enter the pore and the remaining liquid will spread \citep{Delbos2010, Ding2012a}. If the droplet falls beside a pore on a wetting substrate, the liquid will be slightly deflected downwards as it spreads over the pore and a splash is created at the outer edge \citep{Roisman2010}. This is a localised splash, which differs from lamella splashing (\textit{e.g.}, prompt or corona splashing {\citep{Rioboo2001, Yarin2006}}).\\

Just like in the previous examples, our interest is in droplet impact on a smooth surface containing a hole. The difference is that we focus on impacts on closed holes (pits) of varying diameter and depth which we experimentally investigate and compare to the behaviour for impacts on (open-ended) pores. A pit will induce significantly different flow patterns through the interaction of the liquid with both the bottom and the air inside it. This gives rise to richer phenomena than in the pore case. Furthermore, we will address the different regimes by varying the distance of impact to the hole in small steps.

We have structured the paper as follows: In \textsection \refs{expsetup}, we introduce the experimental setup in detail. In \textsection \refs{pits} and \textsection \refs{pore}, we describe our experimental observations for impact near pits and pores respectively. In addition, \textsection \refs{pore} provides a comparison between impact near pit and pore. Finally, \textsection \refs{the_end} summarises our findings.

\section{Experimental details}
 \label{sec:expsetup}
To study the effect of droplet impact near a hole, we use the experimental setup illustrated in figure \reff{setup}\textit{a}. Droplets are generated by expelling liquid from a syringe at a low rate ($\leq 0.4$ ml/min, syringe pump Harvard phd $2000$), through a pipe and into a capillary needle (outer diameter 0.85 mm). The droplet formed at the needle’s tip detaches as soon as the gravitational force overcomes that of surface tension. We use milli-Q water (density $\rho = 998$ kg/m$^3$, surface tension $\sigma = 73$ mN/m and viscosity $\mu =1.0$ mPa$\cdot$s) with about $0.5\%$ in volume of red food dye added to increase contrast (without changing the surface tension). With these properties and the fixed needle size, the droplet diameter equals $D_d \approx 3.0$ mm. Unless stated otherwise, the impact velocity $U_i \approx 3.0$~m/s and therefore the Weber number is given by $\mbox{\textit{We}} = \frac{\rho D_d {U_i}^2}{\sigma} \approx 370$ and the Reynolds number is $\mbox{\textit{Re}} = \frac{\rho D_d {U_i}}{\mu} \approx 9 \cdot 10^3$. Hence, all experiments are performed in the inertial regime and no prompt or corona splashing occurs (the threshold velocity for corona splashing is $4$\---$5$ m/s for water, $\mbox{\textit{We}} = 650$\---$1050$ \citep{Riboux2014}).
 
The target surface is a perspex plate containing a row of six equidistant holes, with a distance between the hole centres of $5.0$ mm which is sufficient to prevent that the flow at one hole affects the flow at the other one. The milli-Q water has a contact angle on the substrate of $70^\circ \pm 10 ^\circ$. The pits and pores were made with a drill which has a tip angle of $118^{\circ}$, such that the bottom of the pit has a slight angle as well (figure \reff{setup}). The edge of the holes is sharp, of about $90^\circ$, and due to the drilling process, the edge can be slightly rough on a scale of $0.1$ mm. For the pit dimensions ({table}~\ref{tab:pitsizes}) we selected a large diameter-to-depth aspect ratio ($2.0/0.5$), a small one ($2.0/4.0$) and one of order unity ($2.0/2.0$). To also assess the importance of the absolute size (or pit to droplet diameter ratios), we selected several different pits, while keeping the aspect ratio fixed at unity ($1.0/1.0$), ($1.5/1.5$) and ($2.0/2.0$). The depth is measured at the pit centre. We tested one pore (\ie, a hole that penetrates through the entire substrate) with a diameter $D_p$ of $2.0$ mm and a length of $10$ mm. 

Two relevant geometrical dimensionless parameters can be denoted for the substrate: $D_d / D_p$ and $D_d / Z_p$. Regarding the first one, if  $D_d / D_p \gg 1$, we speculate that the spreading droplet will be hardly affected by the pit. It will act as small scale roughness, and can enhance prompt splashing \citep{Rioboo2001, Xu2007b}. For $D_d / Dp \ll 1$, the droplet will just spread inside the pit and if impact energy is sufficient, splash at the edges of it (see \cite{Subramani2007}). Therefore, we chose $D_d / D_p$ in this work always slightly larger than $1$, \ie, $D_d \geq D_p$. The other parameter $D_d / Z_p$ varies between $0.75$\---$6$ (pit) and for the pore case, it goes effectively to $0$. This parameter is of interest as it gives the ratio of droplet volume to the volume of air enclosed in the pit and therefore, indicates the relative importance of the role of air.

The key parameter varied in this study, $\Delta$, is given by the distance between the outer pit edge and the nearest drop edge, see figure \reff{setup}\textit{b}. To explore $\Delta$ in a uniform way for the various pit diameters, we normalise it by the pit diameter $D_p$ and denote this dimensionless distance as $\del$:

\begin{equation}
\del = \frac{\Delta}{D_p} = \frac{r_{dp} + D_p/2 - D_d/2}{D_p}  
\nonumber
\end{equation}

where $r_{dp}$ is the distance between the centres of droplet and pit. Figures \reff{setup}\textit{b-d} provide a schematic explanation of the significance of this quantity $\del$. The region $0<\del<1$ corresponds to a partially covered pit at the moment of droplet impact. When $\del$ is greater than 1, the droplet falls besides the pit. At a negative $\del$, the droplet completely covers the pit. Hence, this dimensionless $\del$ contains information about both inner and outer pit edges, which is relevant for the phenomena we observe. This information would be lost if $\Delta$ would be scaled with the droplet diameter. In our experiments, $\del$ is adjusted between $-0.4$ and $1.3$ with small increments. 

The impact events are recorded simultaneously by high-speed cameras from the side (Photron SA$1.1$) and from below (pits: Photron APX-RS, pores: Photron SAX2) at a frame rate of $5-10$ kHz (figure \reff{setup}\textit{a}). Additional high-speed imaging experiments are performed by zooming in from below with a long-distance microscope to capture the flow behaviour inside the pit (Photron SA$1.1$). The droplet diameter $D_d $, impact velocity $U_i$ and the distance between the location of impact and the pit/pore centre, $r_{dp}$, are determined from the recordings.

\begin{figure*}\begin{center}
	\includegraphics[width=1.0\textwidth]{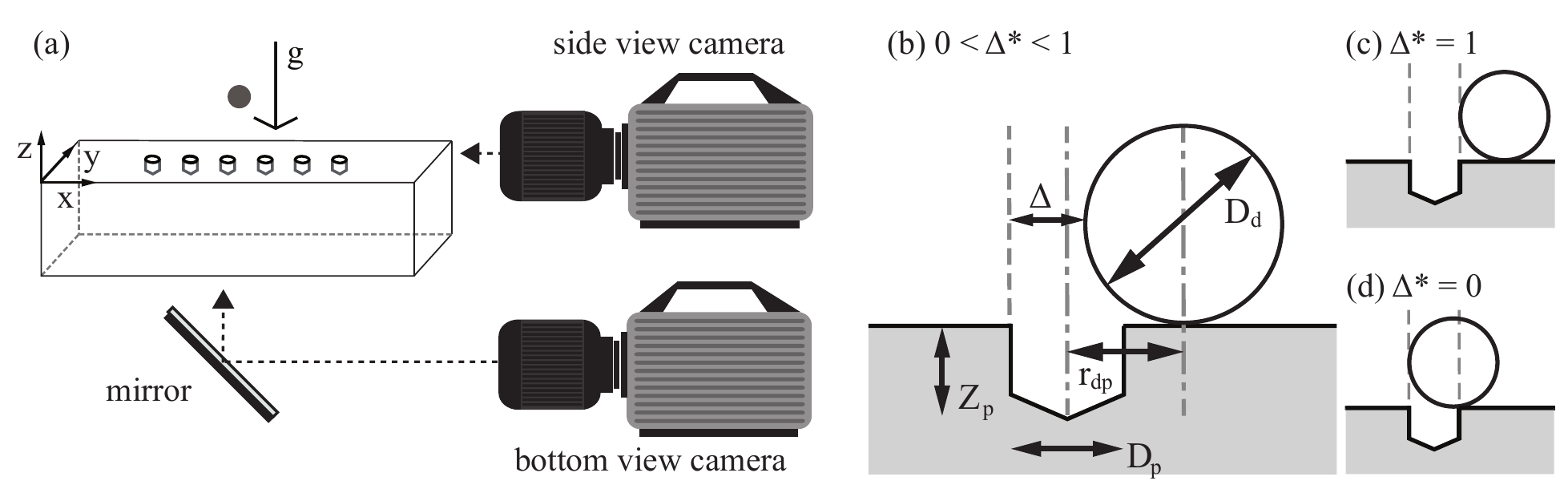} 
	\caption{(\textit{a}) Schematic representation of the impact of a droplet on a plate with a row of holes, in this case closed pits. The impact is recorded simultaneously by both a side view camera and a bottom view camera with help of a mirror. (\textit{b}) The relevant parameters are denoted: droplet diameter $D_d$, pit diameter $D_p$, pit depth $Z_p$, the distance between the centres of pit and droplet, $r_{dp}$, and the distance between the outer pit edge and the nearest droplet edge, $\Delta$. The latter can be normalised with $D_p$ to a new parameter $\del$. All parameters are the same for a pore. (\textit{c}) No overlap occurs when $\del \geq 1$. (\textit{d}) There is complete overlap at $\del = 0$.
	}
	\label{fig:setup}
\end{center}\end{figure*}

\begin{table}
   \centering
   \begin{tabular}{r c c c c c c }
   	   Diameter $ D_p $  [mm]   &     $\ \ 2.0$ & $\ 2.0$ & $\ 2.0$ & $\ 1.0$ & $\ 1.5 $\\
   	   Depth      $ Z_p $  [mm]   &     $\ \ 0.5$ & $\ 2.0$ & $\ 4.0$ & $\ 1.0$ & $\ 1.5 $\\
   \end{tabular}
   \caption{Overview of the investigated pit sizes.}
      \label{tab:pitsizes}
\end{table}

\section{Impact near pits}
	\label{sec:pits}
\subsection{Observed phenomena}
	\label{subsec:pitphen}
%
\begin{table}
   \centering
\begin{tabular}{l l p{9 cm}}
Phenomenon & Hole type& Description \\
\hline
Jet 		& Pit 	& A jet emerges at the inner edge, \ie, the edge closest to the point of impact, and originates 
from liquid that enters the pit and is deflected upwards by the presence of a bottom (and side walls). In time, a jet always occurs after a splash.\\
Air bubble & Pit 	& An air bubble appears for almost full overlap between droplet and pit caused by a pressure build-up inside the latter.\\
Splash 	& Pit or pore & A splash is created as spreading liquid hits the outer edge of pit or pore, which acts as an obstacle to the flow. It is observed for all pit experiments (as long as $\del<1.5$) and always happens before a jet or an air bubble. \\
Slow splash & Pore 	& A slow splash occurs for $1.0 < \del < 1.5$ and has most likely the same origin as the pit splash. It is broad and moves inwards in the horizontal direction. \\
Thin splash & Pore 	& A thin splash also appears at the outer edge, but for $0.5<\del<0.7$. It has less momentum than the pit splash (under the same conditions), is thin and shoots upwards. Notice that for $0.7<\del<0.85$, there is no splash. This marks the distinction between the slow and the thin splash.\\ 
Blob 		& Pore 	& A blob is a slow splash, however, only visible for a few frames and with a negligible velocity. It appears in the region $0.85<\del<1.0$.\\
	\end{tabular}
     \caption{Overview and definitions of various phenomena.}
      \label{tab:phen_explained}
\end{table}

\begin{figure*}\begin{center}
	\includegraphics[width=1.0\textwidth]{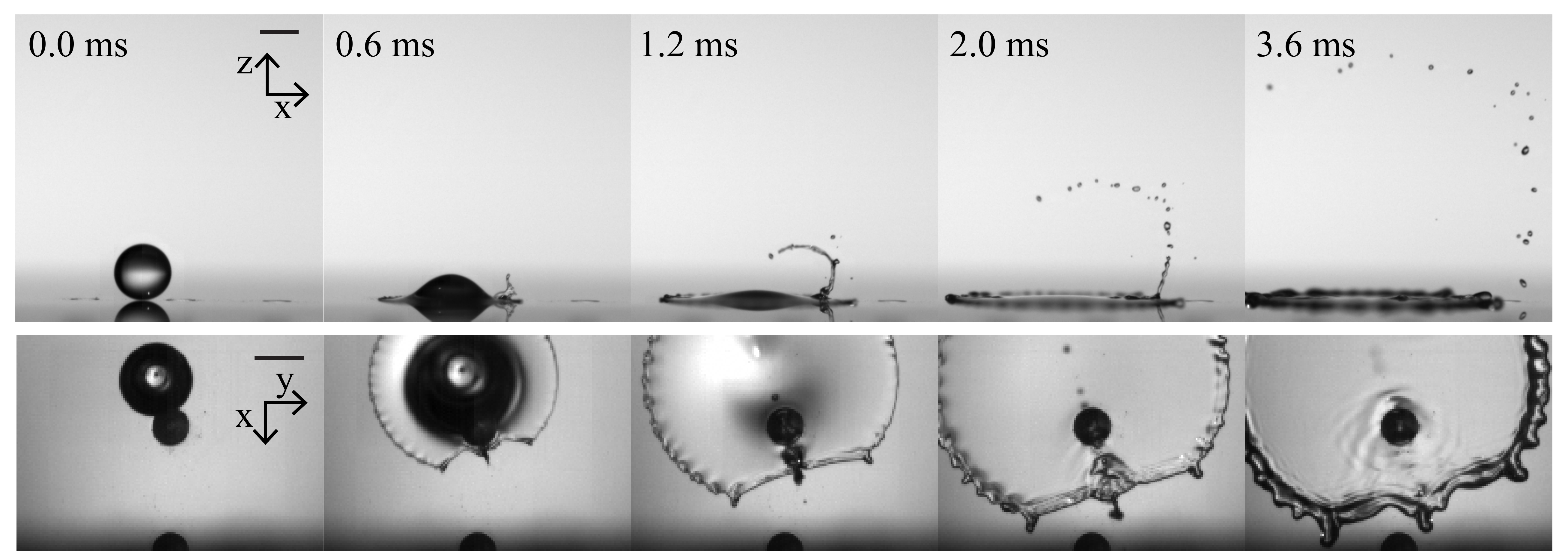}
	\caption{Time evolution of a splash imaged from the side (top) and from below (bottom). The droplet impacts at a dimensionless distance $\del = 0.89$ from a pit with a diameter $D_p = 1.5$~mm and a depth $Z_p = 1.5$ mm. At $t = 0.6$ ms the splash is created by the deflection of liquid at the outer pit wall. The splash velocity is $4.8$ m/s, averaged over the first millisecond. Droplet diameter $D_d$ and velocity $U_i$ are $2.9$ mm and $2.93$ m/s respectively. The bar in the two leftmost figures represents $2$ mm. Note that the orientation of the side and bottom view differs by $90^{\circ}$, see figure \reff{setup}a for the coordinate system.
	 }
	\label{fig:phen1}
\end{center}\end{figure*}

\begin{figure*}\begin{center}
	\includegraphics[width=1.0\textwidth]{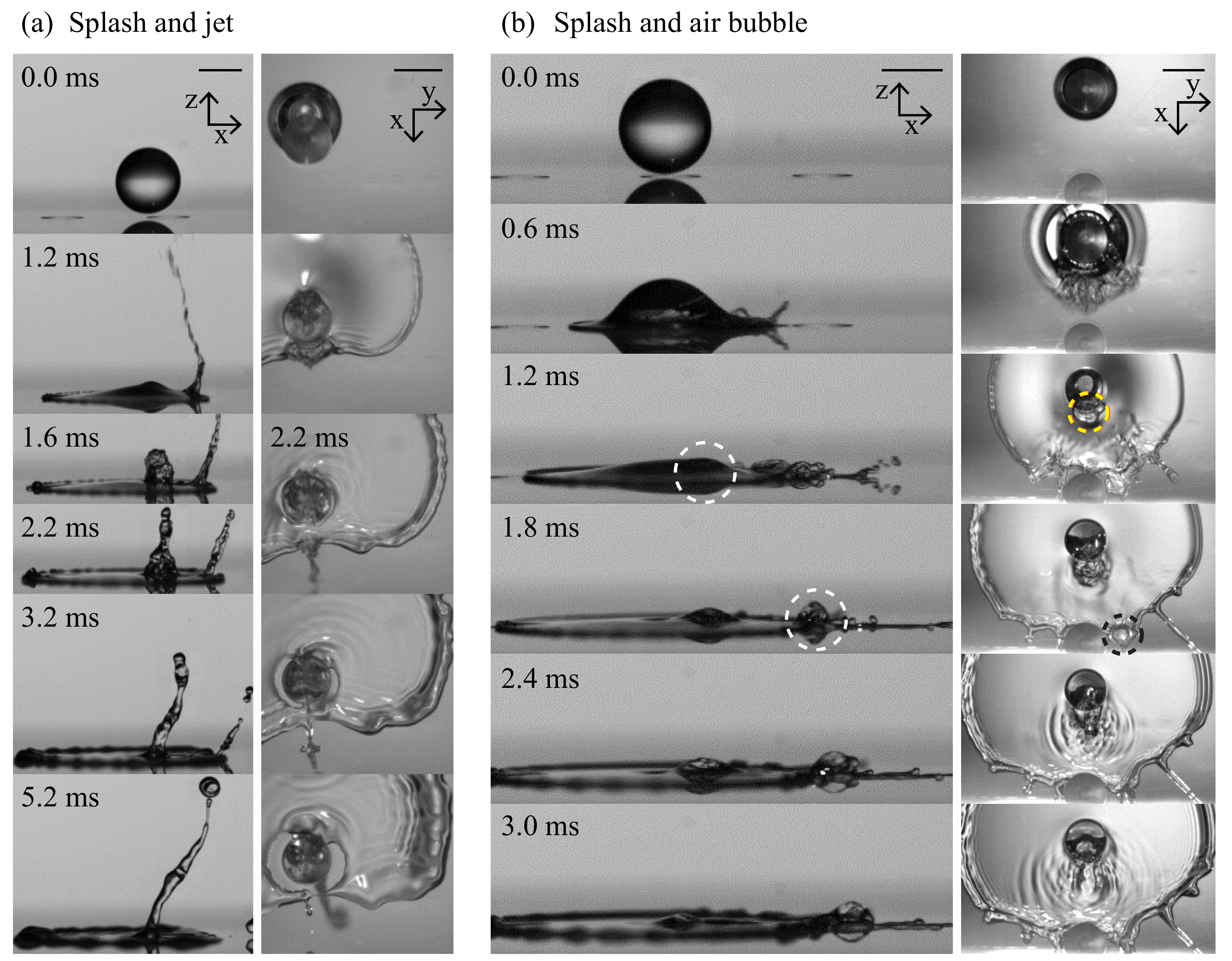}
	\caption{Snapshots of two distinct phenomena: The formation of a jet (\textit{a}) and the appearance of an air bubble (\textit{b}) imaged from the side (left) and from below (right). The bar in the top figures represents $2$ mm. Note that the orientation in side and bottom view differs by $90^{\circ}$. For both cases the parameters are as follows: $D_p = 2.0$ mm, $Z_p = 2.0$ mm, $D_d = 3.1$ mm, $U_i = 2.99$~m/s. 
	(\textit{a}) For a dimensionless distance of $\del = 0.16$, first a fast splash appears at $t=1.2$ ms, and subsequently a jet is observed at $t=1.6$ ms. The velocities in the first millisecond are $17.7$ m/s for the splash and $2.8$ m/s for the jet.
	(\textit{b}) For full overlap between droplet and pit ($\del = -0.25$), the resulting phenomenon is totally different. Air is pushed out of the pit by the impacting water. An air bubble is visible at the pit edge at time $t=1.2$ ms and another air bubble is observed at the spreading edge from $t = 1.8$ ms onwards. The first is the air bubble we will focus on in this study. Both are indicated with dashed circles.
	 }
	\label{fig:phen23}
\end{center}\end{figure*}

Droplet impact near a pit may result in any of the following three phenomena: (1) a splash, where spreading liquid is partly diverted at the outer pit edge into the vertical direction; (2) a jet, where some liquid is expelled from the pit at the inner pit edge in an almost exclusively vertical direction; and (3) an air bubble, which is seen to emerge from the pit. More details on the phenomena are given in table \ref{tab:phen_explained}. Figures \reff{phen1} and \reff{phen23} present a series of side and bottom view images extracted from the recordings, see movie $1$ in the supplementary material. 

\subsubsection{Splash}
In figures \reff{phen1} and \reff{phen23}, we can see examples of splashes forming as spreading liquid impacts on the outer edge of the pit, with which we denote the spot that is most distant from the droplet's impact location. Although this phenomenon is observed for the whole range of explored $\del$, the direction and velocity of the splash varies. For small overlap (figure \reff{phen1}), the splash is thin, curved inwards and its speed is comparable to the impact velocity. As overlap increases (figure \reff{phen23}\textit{a}), the splash becomes faster and more vertical. For full overlap (figure \reff{phen23}\textit{b}), it is broad, slow and moving mainly horizontally. The splash velocity is discussed in detail in \textsection \refs{splashvel}. From the high-speed imaging experiments capturing the flow inside the pit (figure~\reff{zoom_pics}\textit{a}), we found that liquid is also deflected downwards and that some will remain in the pit after the impact event. The flow is captured in a simple schematic in figure \reff{jetflow}\textit{a}.

\begin{figure*}\begin{center}
	\includegraphics[width=0.70\textwidth]{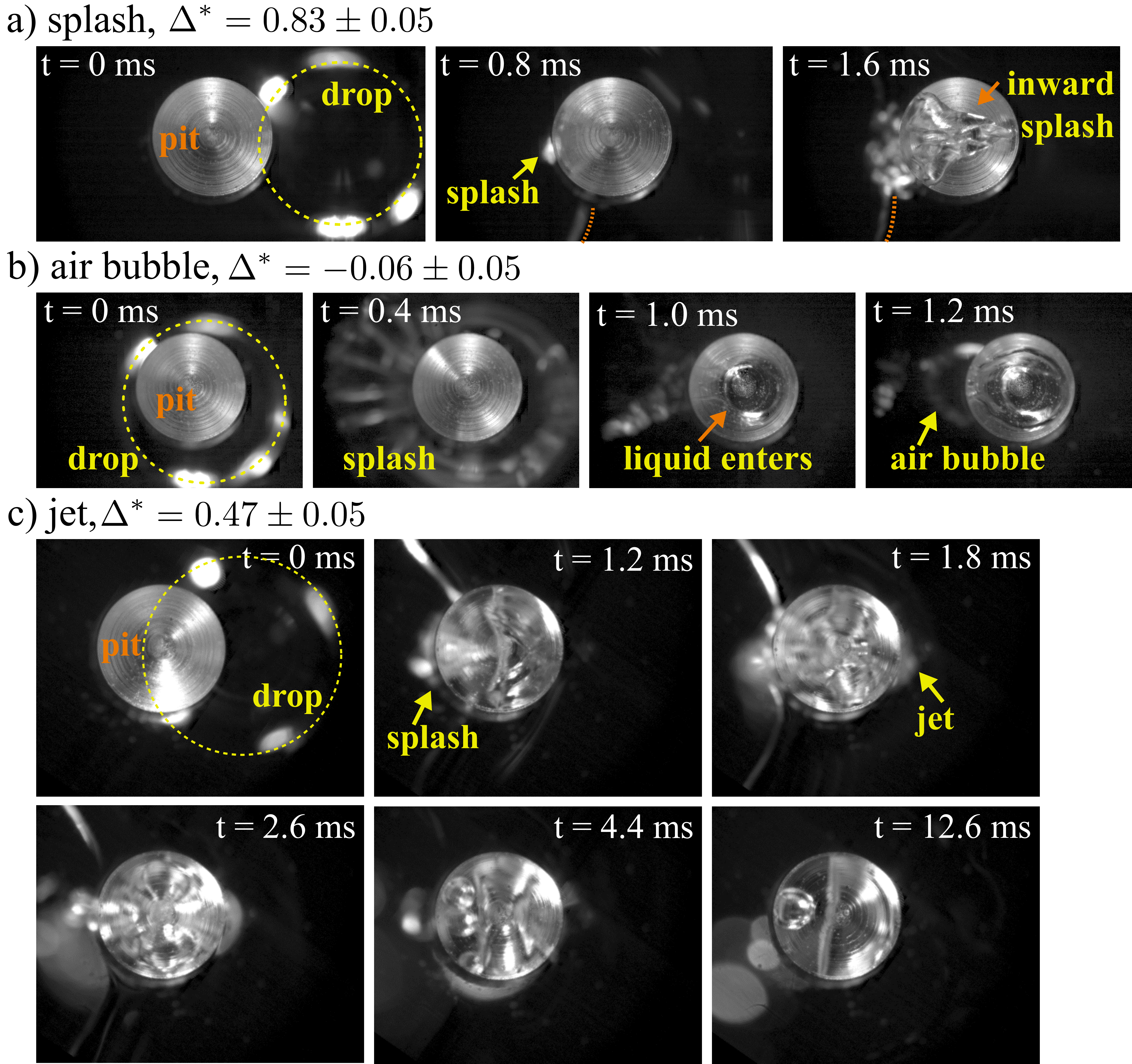}
	\caption{Images showing the flow inside a pit of $2.0$ mm in both diameter and depth. (\textit{a}) For small overlap between droplet and pit, liquid enters the pit at the outer edge, resulting in an inward flow. (\textit{b}) For full overlap, after liquid enters the pit, an air bubble emerges.(\textit{c}) For partial overlap between drop and pit, the flow behaviour is more complex, as liquid and air mix. However, a splash can be seen at the outer edge and a jet at the inner one (as can also be observed in figures \reff{phen23}\textit{a} and \reff{flowinpit}).
	}
	\label{fig:zoom_pics}
\end{center}\end{figure*}

\begin{figure*}\begin{center}
	\includegraphics[width=0.9\textwidth]{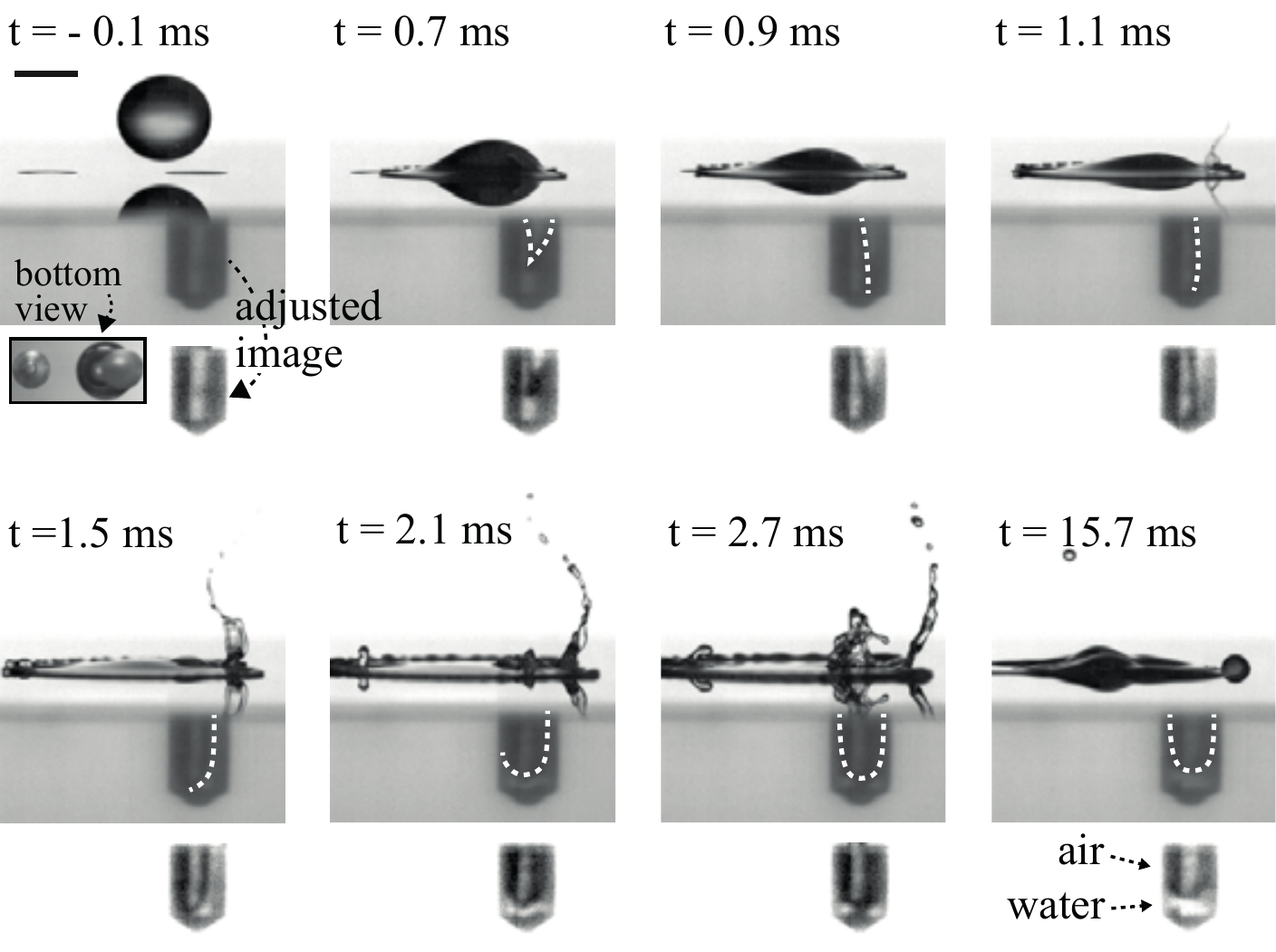}
	\caption{For partial overlap ($\del = 0.22$), the liquid enters at the outer side of the pit, is deflected upwards by the bottom, after which a jet appears at the inner edge. Each frame shows the spreading droplet on top of the substrate as well as the flow inside the pit, with an extra picture for the flow inside the pit with enhanced brightness and contrast directly below it. Pit diameter and depth are $2.0$ and $4.0$ mm respectively (The leftmost pit is shallow in depth, and hence not visible from the side). The inset in the first figure gives the bottom view at the moment of impact, showing that the droplet impacts in line with the pits and hence in the plane of the side view camera. The scale bar represents $2.0$~mm. Other parameters are as follows: $D_d = 3.0$~mm, $U_i = 3.03$~m/s and $U_s = 9.2$~m/s. }
	\label{fig:flowinpit}
\end{center}\end{figure*}

\begin{figure*}\begin{center}
	\includegraphics[width=0.75\textwidth]{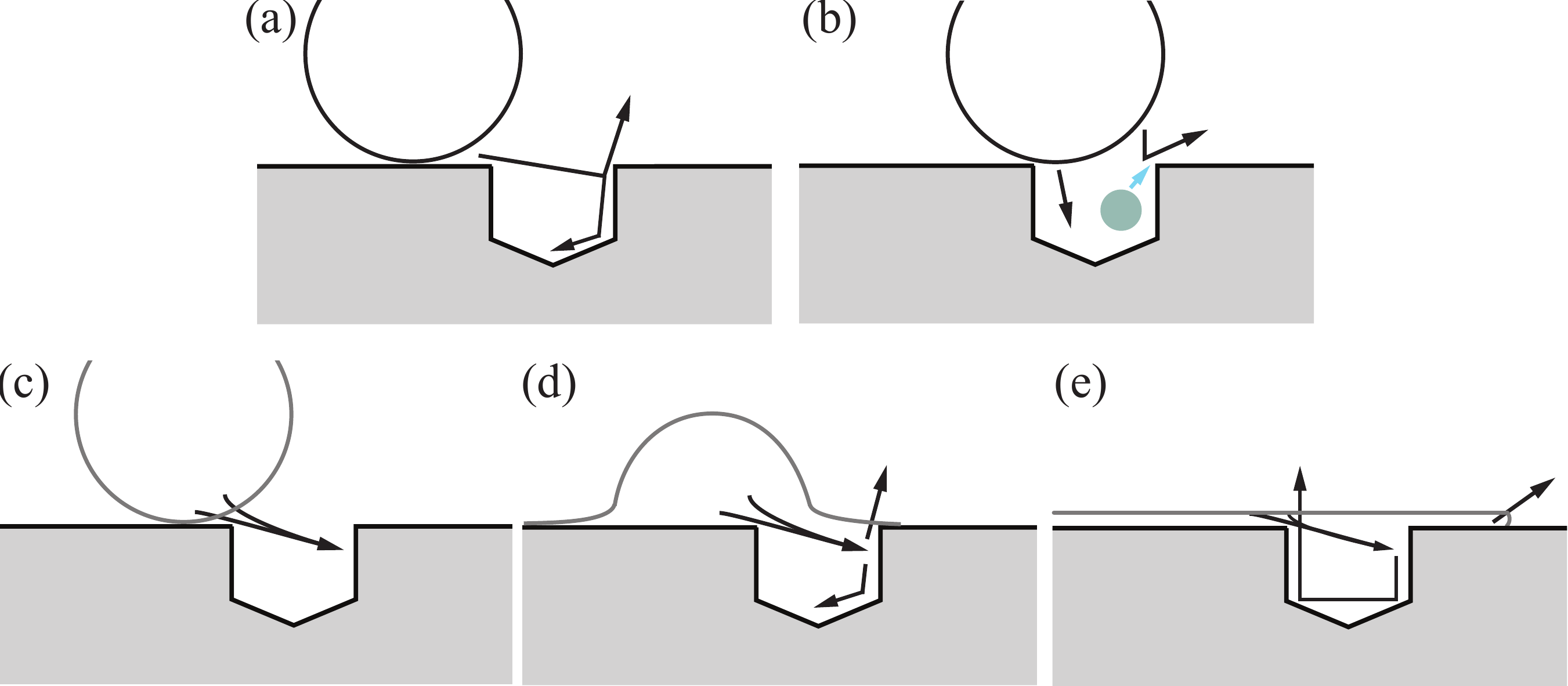}
	\caption{ A schematic representation of the (possible) flow shortly after impact as inferred from high-speed images taken from below with a long-distance microscope focussing on the pit (see figure \reff{zoom_pics}). (\textit{a}) For little overlap, liquid spreads outwards, hits the outer pit edge and is deflected both upward and downward. (\textit{b}) For full overlap, liquid flow is separated: a broad splash outwards (right arrow) and liquid entering the pit near the inner pit edge. This liquid pushes out an air bubble at the outer pit edge (filled circle).  (\textit{c-e}) Liquid hits the outer pit edge (\textit{c}), creating an upward and downward flow (\textit{d}), resulting in a splash at the outer edge and a jet at the inner one, penetrating the flat drop (\textit{e}).
	} 
	\label{fig:jetflow}
\end{center}\end{figure*}

\subsubsection{Jet}
A jet is only created when the impacting droplet overlaps with the pit. The necessary amount of overlap depends on the pit depth. In figure \reff{phen23}\textit{a} ($\del = 0.16$) two phenomena occur: first, a splash at the outer edge (at $t=1.2$ ms) and subsequently, a thicker object appears at the inner edge, which we denote as a jet (see also supplementary movie $1$). In this case, there are two competing flow mechanisms: on the one hand, the spreading liquid that leads to splashing as described above; on the other hand, as there is overlap between droplet and pit, part of the liquid can immediately enter.

{By what mechanism is the jet created? The flow inside the pit cannot be directly extracted from the zoomed-in bottom views (figure \reff{zoom_pics}\textit{c}). For pits of $4.0$~mm in depth, however, the flow is visible on the side view images, see figure \reff{flowinpit} and supplementary movie $2$. As the pit is not totally in focus, the air-water interface is indicated with a dashed line. The extracted flow, which is also illustrated in figure \reff{jetflow}\textit{c-e}, is as follows: after impact, the liquid will spread outwards and hit the outer edge, resulting in an upward and a downward splash. The latter will continue on the pit bottom and will be deflected upward at the inner edge, resulting in a jet. The location where the jet appears, indicates that the outward flow dominates the downwards one. Such flow path is consistent with the observed jet emergence times, which will be discussed in \textsection \ref{subsec:jettime}. 
}

\subsubsection{Air bubble}
 \label{sec:airbub}
Another observed phenomenon, an air bubble, is found for large overlap ($\del \approx 0$). Here we show the case for which $\del = -0.25$ (figure \reff{phen23}\textit{b} and in supplementary movie 1 for $\del=-0.14$). The air bubble is visible at time $t = 1.2$~ms at the outer edge. Subsequently, it moves slightly outwards ($t = 1.8$ ms) and breaks-up ($t = 3.0$~ms). From the bottom view images (figure \reff{zoom_pics}\textit{b}), we can see that the central part of the droplet enters the pit. The corresponding flow is sketched in figure \reff{jetflow}\textit{b}.

We state that the air bubble is pushed out as a consequence of liquid sealing the pit. The driving mechanism is a pressure build-up due to liquid entering the closed-off pit, \ie, the air inside it gets compressed. There are three pressures to be considered: the dynamic impact pressure, the hydrostatic pressure and the Laplace pressure. The dynamic pressure is relevant at early times, also because it is responsible for liquid entering the pit. Its initial value scales with $\rho U_i^2 \propto 10$~kPa and is observed (from simulations) to drop quickly after $\tau = D_d / (2 U_i)$  for droplet impact on both a liquid film and a solid substrate \citep{Josserand2003, Eggers2010}: it is one order of magnitude smaller at $t \approx 1.5 \tau$ and two orders of magnitude at $t \approx 3.0 \tau$. In our case, the impact time scale is $\tau = 0.5$~ms. The hydrostatic pressure applied by the liquid film is $\Delta P_{film} = \rho g d$, with $g$, the gravitational acceleration and $d$, the thickness of the film. Even if the liquid film were $1$ mm thick, the film pressure $\Delta P_{film} $ would be approximately $10$ Pa and would therefore constitute a negligible obstacle for the air bubble to move upwards.  Calculating the Laplace pressure, one finds it to be in the same order of magnitude as the pressure build-up inside the pit $\Delta P_{pit}$. Therefore, it will influence the final shape and size of the bubble, but will not prevent it from appearing.

Hence, the air has to get compressed, such that the pressure build-up $\Delta P$ inside the pit overcomes the impact pressure. With Boyle's law we can estimate the volume of compression that is needed: $\Delta V = V_p \Delta P / (P_0 + \Delta P) \equiv  \alpha V_p $. The precise value of $\Delta P$ depends on the moment the pits gets sealed, \ie, on the droplet velocity at that time. As we show in figure \reff{phasediagram}, air bubbles are observed for large overlap only, and consequently, time of pit closure is rather short. Therefore, using $U_i$ is a reasonable estimate: $\Delta P \approx \tfrac{1}{2} \rho U_i^2$ and accordingly $\Delta V \approx 0.04 V_p$. This indicates that only a small amount of liquid needs to enter the sealed pit, in order to reach a pressure inside it opposing the dynamic pressure.

As the dynamic pressure quickly drops after $t> \tau$, one can expect air bubbles to form slightly after. We observe air bubbles in a time range of $0.8 - 1.6$~ms. Hence, as soon as the liquid is decelerated sufficiently, the pressure build-up inside the pit becomes dominant and thus the appearance of the air bubble is directly related to the amount of liquid that enters the pit: for all practical purposes the air within the pit can be considered to be incompressible. 

For almost the whole $\del$ range, there is an asymmetry between the liquid entering the pit and the pit itself, therefore designating where the air bubble appears. However, when the centres of droplet and pit are aligned at impact ($r_{dp} = 0$), a Rayleigh-Taylor (or Richtmyer-Meshkov) instability can play a role. Its characteristic acceleration is given by $a^* = U_i^2 / (D_d/2 + \alpha Z_p)$, where $\alpha$ is the pressure fraction calculated above: $\alpha=0.04$, indicating that the air quickly decelerates the liquid. The unstable wavelengths for water are $\lambda_{unstable} > 2 \pi \sqrt{{\sigma}/({\rho a^*})} \gtrsim 0.7$~mm for all the investigated pit depths. Therefore, this instability can dominate for head-on impact as it is smaller than the pit diameters, which vary from $1.0 - 2.0$~mm.

\subsection{Phase diagram}
\begin{figure*}\begin{center}
	\includegraphics[width=1.0\textwidth]{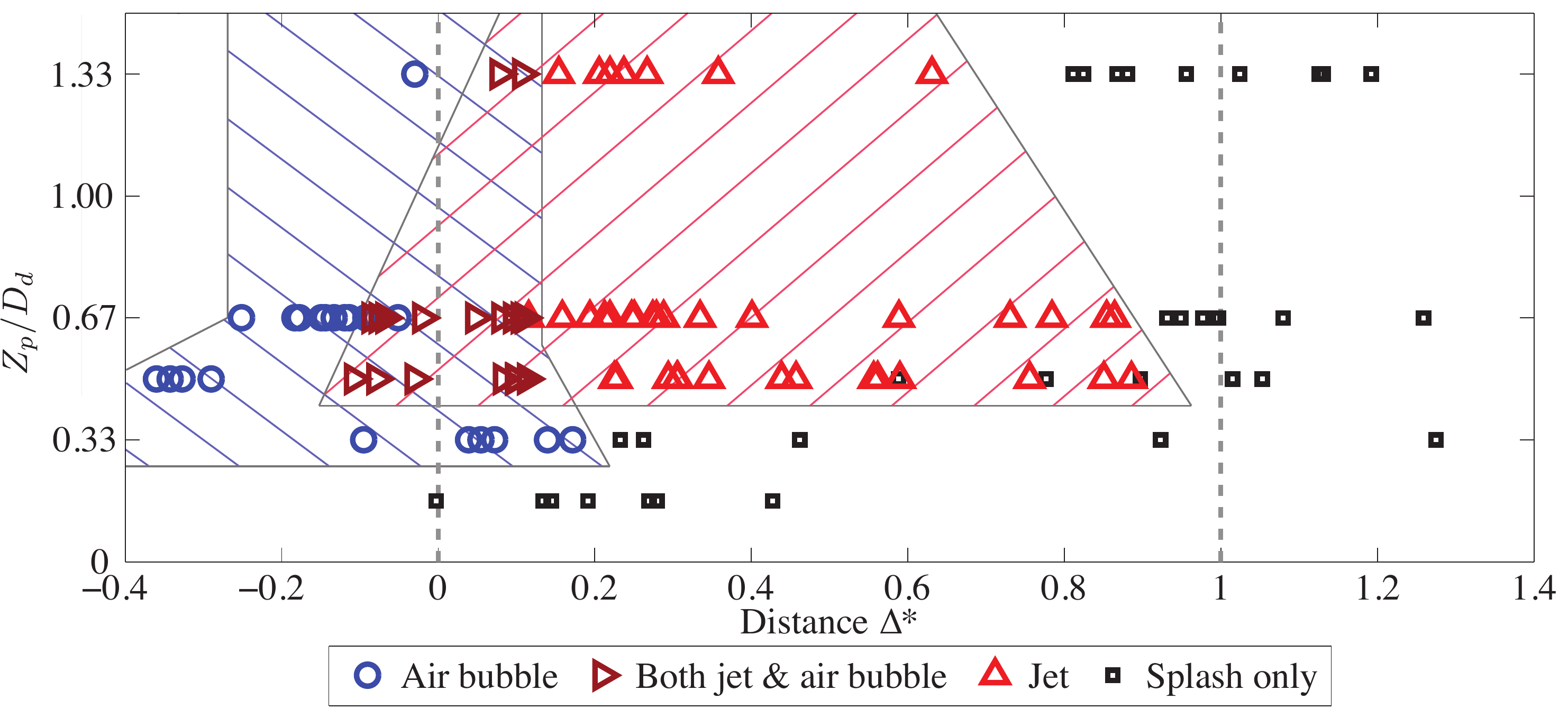}
	\caption{Phase diagram of the three different types of phenomena (splash, jet and air bubble) in the parameter space spanned by the dimensionless distance between the outer pit edge and nearest drop edge $\del $ and the dimensionless pit depth $Z_p/D_d$. All experiments are done in the inertial regime for a Weber number of $\mbox{\textit{We}} \approx 370$. The shaded regions and their boundaries are guides for the eye: the left one (~$\backslash \ \backslash$~) represents the region where an air bubble appears, the right shaded area (~/ /~) shows the region where a jet is present. The droplet diameter and velocity are kept constant within a $3$\% accuracy to $3.0$ mm and $2.95$ m/s, respectively. Note that a splash is observed in all cases, albeit sometimes small.
	}
	\label{fig:phasediagram}
\end{center}\end{figure*} 

All of the phenomena discussed above are plotted in the phase diagram of figure \reff{phasediagram}, which maps their dependence on the dimensionless distance $\del$ and the dimensionless pit depth $Z_p/D_d$. The broad picture is as follows: Decreasing $\del$ from large to intermediate to small, we move from a regime with a splash only, through one with a splash and a jet, into a region with a splash together with an air bubble. Decreasing the dimensionless pit depth enlarges the range (of $\del$) for which a jet is observed, whereas the domain in which an air bubble is found remains approximately constant in size until the pit becomes too shallow to produce an observable bubble or jet. 

The transition between splash only and jetting (near $\del = 1$) suggests that for deeper pits a larger overlap is needed to produce a jet. This is probably related to the longer travel path ($2 Z_p + D_p$) before the liquid can exit a deeper pit (\ie, as a jet; figure~\reff{jetflow}\textit{c-e}). There are two lines of reasoning regarding the relation between depth and jet formation. First, a long travel path has more dissipation than a short one, and second, as a longer path will take a longer time, there can be a larger droplet mass above the liquid in the pit, preventing it from leaving. On the other side of the regime, near $\del = 0$, where jetting gives way to the formation of a bubble only, the reasoning is similar. Again, we think that the disappearance of the jet, that occurs for larger $\del$ as the pit depth increases, is related to the path that the liquid needs to travel. In addition, for full overlap ($\del < 0$), a jet is noticed at the outer edge, appearing after the splash.

The air bubble regime is bound between $\del \approx 0.1$ and the minimum value of $\del$, where the latter by its definition increases with increasing pit diameter, causing the blank region in the top left of figure \reff{phasediagram}, where no measurements are possible for the pits we have used. We found that the location of the first-mentioned boundary (around $\del \approx 0.1$) solely depends on the absolute distance $\Delta$, \ie, the transition is invariant of the pit diameter and depth. For all pit sizes tested, the transition occurs at $\Delta = 0.22$ mm. We believe that the invariance for the pit diameter $D_p$ relates to the spreading front of the droplet nearby this critical $\Delta$. The spreading liquid overcomes the disruption by the pit for this small distance and encloses the pit quickly from both sides. In addition, it is independent of pit depth, because the pressure build-up inside the pit is sufficient for any of the pit depths (see end of \textsection \refss{pitphen} for details); it suffices to have a trapped amount of air. Note that the jet regime and air bubble regime partly coincide ($\del \approx 0.1$). It means that there is a slow transition from mainly outward spreading liquid (figure \reff{jetflow}\textit{c-e}) to liquid that mainly enters the pit (figure \reff{jetflow}\textit{b}).  

Finally, we speculate the changes to the phase diagram of figure \reff{phasediagram} if we would increase the drop size.  Keeping $\del$ constant, this will increase the mass flow towards and inside the pit, and therefore it can be expected that the regime where jets are observed will be slightly extended towards higher values of $\del$. We do not anticipate a similar change for the air bubble regime, since the liquid volume entering the pit does not depend on the droplet size (cf.  \textsection \refs{airbub}).

\subsection{Splash velocity \label{sec:splashvel}}

Figure \reff{velsplash}\textit{a} shows a plot of the splash velocity against $\del$ for various pit sizes. First, we can observe that for large distance ($\del \gtrsim 1$) the splash velocity is close to the impact velocity. Second, the splash velocity increases rapidly with decreasing $\del$, until a maximum is reached of seven times the impact velocity at $\del \approx 0.2$. Third, beyond that maximum, the splash velocity $U_s$ decreases with decreasing distance. The splash velocity was measured always within $1.0$~ms from its first appearance and was obtained by averaging over $1.0$ ms to diminish effects of shape fluctuations of the tip droplet. In very few cases, the averaging was done for a shorter time period, typically when splash velocities were small. For fast splashing, air drag will have an effect on the droplets, as the droplets are far above their terminal velocity, especially when they are small \citep{Thoroddsen2012a}. However, this velocity reduction in the averaging time is within our measurement error.
Figure \reff{velsplash}\textit{b}, for a pit of $2.0$ mm in both diameter and depth and different impact velocities, indicates that the splash velocity can be normalised with the impact velocity. Some authors have proposed that $U_s \sim U_i^{3/2}$ for another type of splashing, namely the appearance of micro-droplets at the first instant of prompt splashing \citep{Thoroddsen2002, Thoroddsen2012a}. Such a scaling does not lead to a better collapse of the data. (For the above, note that this is the only instance in this study where the impact velocity is varied.)
\begin{figure*}\begin{center}
	\includegraphics[width=0.8\textwidth]{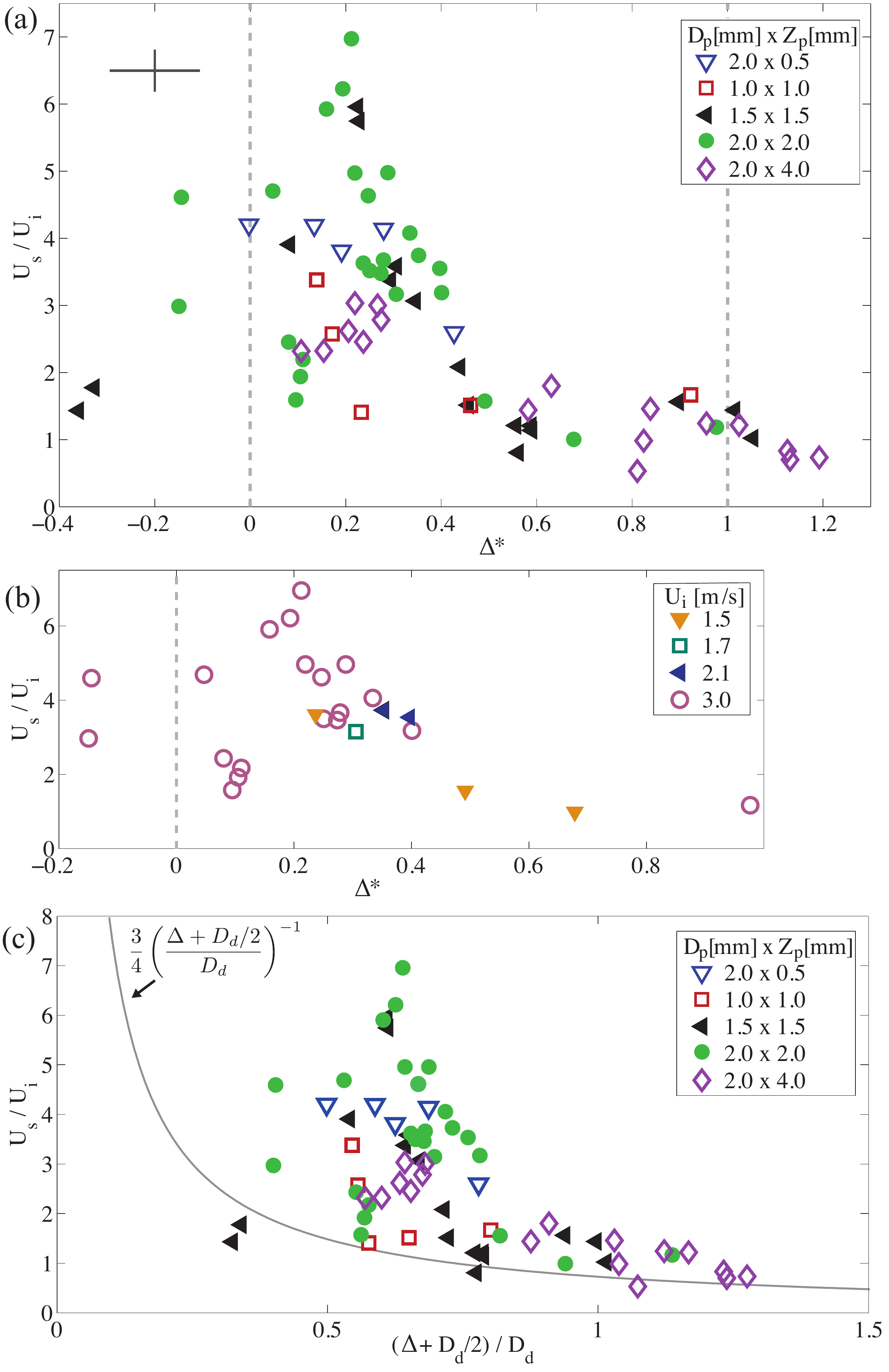} 
	\caption{(\textit{a}) The dimensionless splash velocity $U_s/ U_i$ as a function of the dimensionless distance $\del$ for various pit sizes. For decreasing distance $\del$ the splash velocity $U_s$ increases up to a maximum of $7U_i$ at around $\del = 0.2$. An average error bar is given at the top left. (\textit{b}) A selection of the data of (\textit{a}), for $2.0$ mm $ \times\ 2.0$ mm pits, with various symbols indicating the different impact velocities. (\textit{c}) The dimensionless splash velocity $U_s/ U_i$ is now shown as function of the dimensionless distance between droplet centre and pit edge, $(\Delta + D_d/2)/D_d$. The solid line is related to the lamella spreading theory of \cite{Riboux2014}. 
			}
	\label{fig:velsplash}
\end{center}\end{figure*}

As we consider the behaviour of the individual pit sizes in figure \reff{velsplash}\textit{a}, we notice that to reach high splash velocities a pit depth of $1.5-2.0$ mm is optimal (filled symbols). For a deeper pit of $4.0$ mm, $U_s$ levels off ($\diamondsuit$), which is also the case for the shallowest pit ($\triangledown$). One outlier is the pit of $1.0$ mm in both diameter and depth ($\square$), because the splash velocity only increases from $\del \leq 0.2$, which is, interestingly, also the critical $\del$ for the transition to air bubbles for this pit size. For all other pits, the upward trend in velocity starts from $\del \approx 0.5$. For each pit size, the data is consistent within its measurement uncertainties, except for a pit depth of $2.0$ mm ($\bullet$) at $\del \approx 0.1$. We believe that the main flow behaviour at this $\del$ is very sensitive to small fluctuations during impact, since this $\del$ domain is a transition region, in which both a jet and an air bubble occur. 

\subsubsection{Splash velocity related to lamella spreading}
For droplet impact on solid substrates, the spreading of the lamella in time is widely studied \citep{Rioboo2002, Mongruel2009, Riboux2014}. A scaling for this spreading radius in time is easily obtained with simple dimensional analysis. However, only recently, by \cite{Riboux2014}, an exact prediction was obtained, which matches well with lamella spreading experiments. They find, by using potential flow theory combined with Wagner's theory \citep{Wagner1932}, that the spreading of the wetted area goes as $R / (D_d/2) = \sqrt{3 t U_i / (D_d /2)}$. This can be related to our experiments by looking at the distance between the droplet centre and the pit edge, which is $L \equiv \Delta + D_d/2$ (see figure \reff{setup}\textit{b}). We can try to relate $L$ to the splash velocity, by first taking its time derivative: $dL/dt =  {\scriptstyle\frac{1}{2}} \sqrt{3/2 D_d U_i /t} =  {\scriptstyle \frac{3}{4}} D_d U_i/ L$. If $U_s$ scales with $dL/dt$, the derivation gives $ U_s/ U_i \sim {\scriptstyle \frac{3}{4}} / (L/D_d)$.  Note however that the theory of \citet{Riboux2014} pertains to the spreading of the circular wetted area and not to the faster ejecta spreading. Hence, a slightly larger prefactor should be expected.

We show the prediction in figure \reff{velsplash}\textit{c}, plotting the dimensionless splash velocity versus $L/D_d = (\Delta + D_d/2)/D_d$. Interestingly, the data follows this scaling for $L/D_d > 0.8$. However, especially for the fast splashes (solid symbols), the data clearly deviates from lamella spreading, suggesting a different mechanism. We think that the relevant difference in experiments with various pit sizes, is the amount of air initially in the pit and the resulting outward air flow as the liquid enters into it.

\subsection{Time until jet formation \label{subsec:jettime}}

The jet is caused by liquid that enters the pit (mainly) at the outer edge and travels a path $s_{\text{jf}} \simeq 2Z_p + D_p$ to appear at the inner pit edge, see figure \reff{jetflow} for a sketch. We measured the time between impact and jet formation, $t_{\text{jf}}$, and show it in figure \reff{jettime}\textit{a} as a function of $\del$. The jet formation time is always found to be longer for larger pits. When it is normalised with ${s_{\text{jf}}}/{U_i}$, the data collapses nicely, see figure \reff{jettime}\textit{b}. Furthermore, we note that the proper velocity to collapse the jet time, $t_{\text{jf}}$, is the impact velocity, rather than the splashing velocity. In the region $0< \del < 0.4$, $t_{\text{jf}}$ hardly changes; in contrast to the splashing velocity (figure \reff{velsplash}\textit{a}), which changes rapidly in this regime. Moreover, the splashing velocity reaches significantly different values when depth is varied.

\begin{figure*}\begin{center}
	\includegraphics[width=1.0\textwidth]{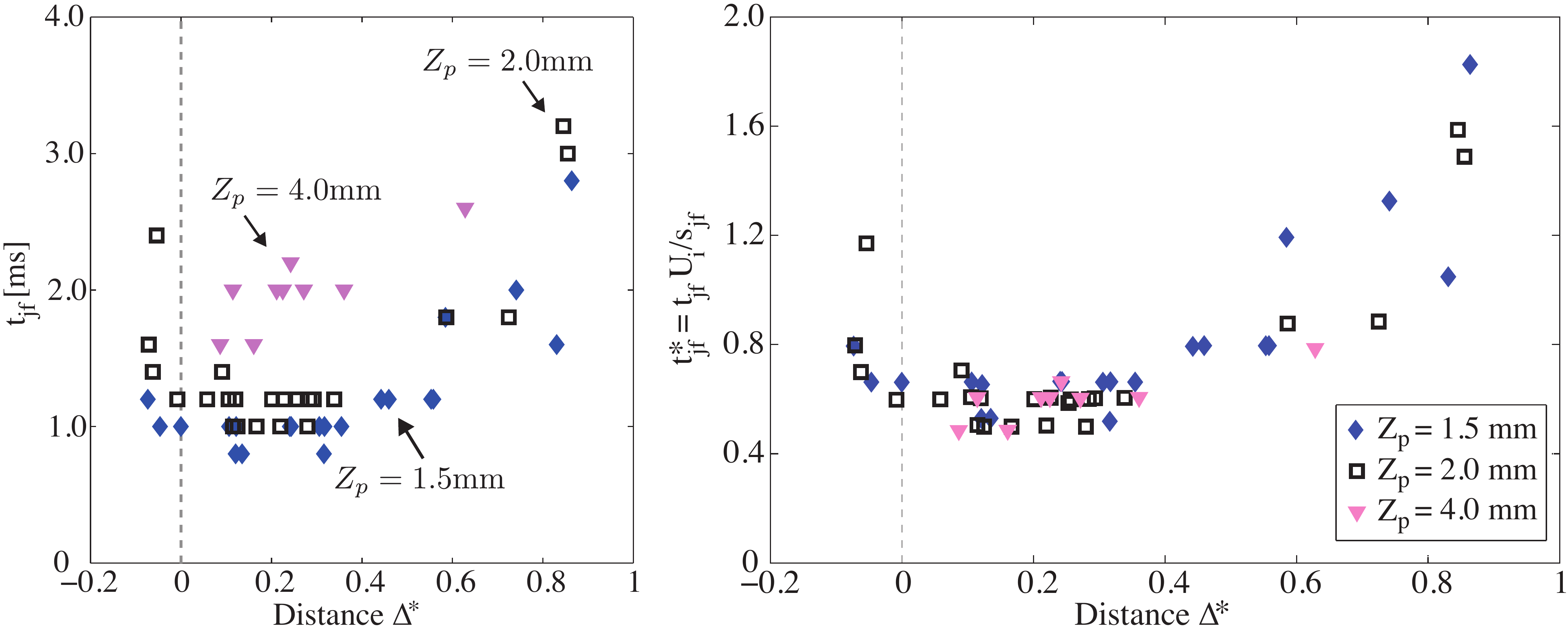}
	\caption{(\textit{a}) The time that it takes for a jet to shoot out from the moment of impact, $t_{\text{jf}}$ as a function of the dimensionless distance $\del$ for three different pit depths $ Z_p$. (\textit{b}) The non-dimensional jet formation time (scaled with the impact velocity $U_i$ and the travel path $s_{\text{jf}}$) as a function of $\del$. The data collapses nicely.
	}
	\label{fig:jettime}
\end{center}\end{figure*}

\section{Impact near a pore} \label{sec:pore}

\subsection{Observed phenomena \label{subsec:phenpore}}

\begin{figure*}\begin{center}
	\includegraphics[width=1.0\textwidth]{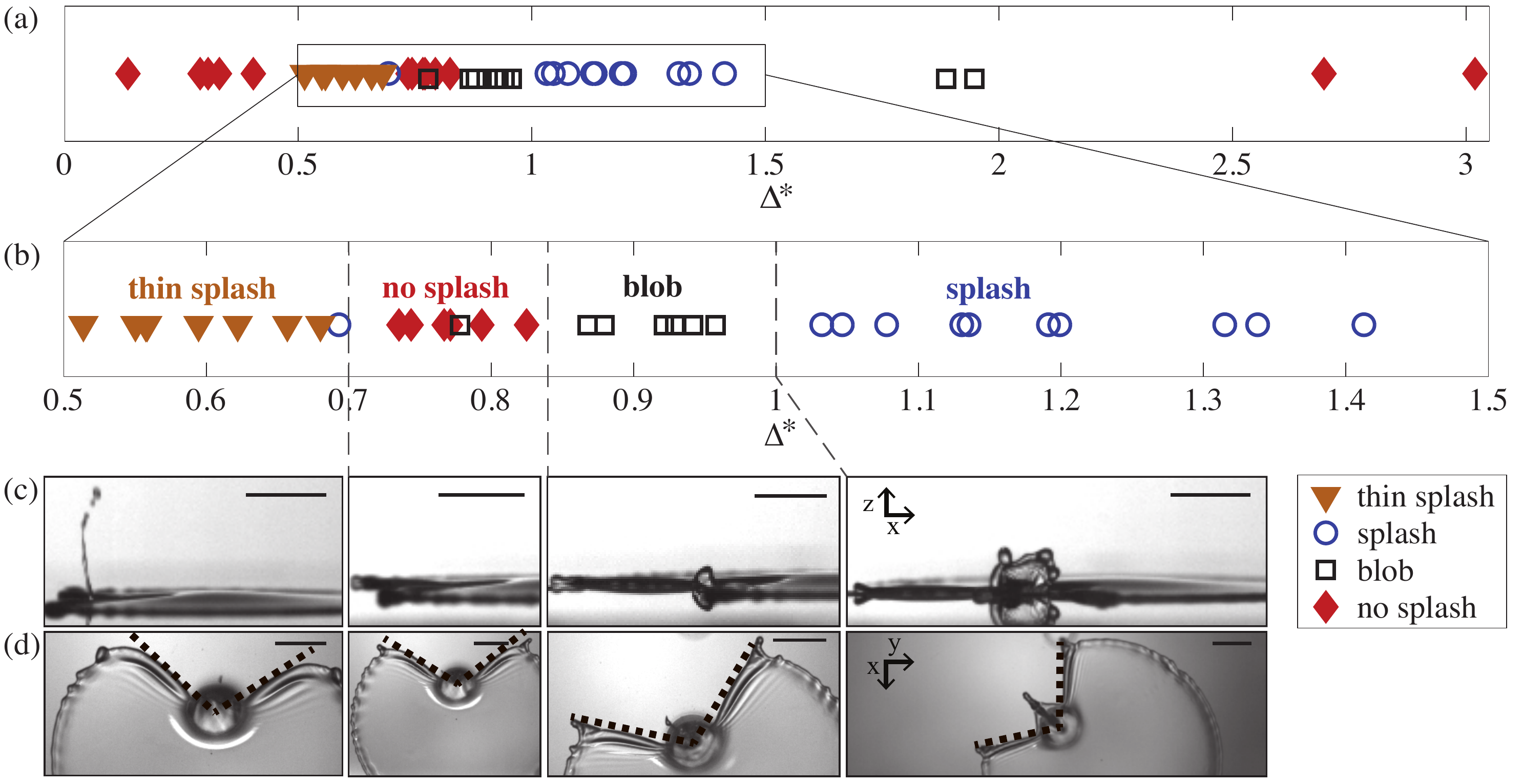}
	\caption{The various phases observed when a droplet impacts near an open-ended pore: a splash ($\bigcirc$), a thin splash ($\triangledown$), a blob ($\Box$) or planar spreading ($\diamondsuit$). (\textit{a}) Phase diagram of different kinds of pore splashes. (\textit{b}) A zoom of (\textit{a}). (\textit{c-d}) Images of each phenomenon, with side view (\textit{c}) and bottom view (\textit{d}).  Snapshots (extracted from supplementary movie $3$) are all taken $1.7$~ms after impact and from left to right $\del$ is given by $0.55, 0.74, 0.94$ and $1.14$ respectively. The bar in each image represents $2.0$~mm. Note that the orientation in side and bottom views differs by $90^{\circ}$. The dashed lines in (\textit{d}) indicate the area near the pore that remains dry throughout spreading. 
	}
	\label{fig:phase_pore}
\end{center}\end{figure*}

Impact near an open-ended pore results in a splash or no splash, \ie, neither jets nor air bubbles are observed. This stands to reason in view of the mechanisms by which the latter two are formed: jet and air bubble need a bottom wall and an enclosed cavity with air for their formation respectively. However, three distinct splash types are observed: a splash, a thin splash and a blob, all caused by spreading liquid hitting the outer pore edge. See table \ref{tab:phen_explained} for the splash definitions. Figure \reff{phase_pore} presents these phenomena as a function of $\del$ along with experimental snapshots (which are also added as supplementary movie $3$). Surprisingly, there are two distinct splashing regimes separated by a region in which planar spreading occurs.

For $1.0 < \del < 1.5$, we observe a slow splash, which descends into a region with a blob at both $\del <1.0$ and $\del >1.5$, until just planar spreading is seen. Note that a blob denotes a splash that is only visible for a short time and has almost no velocity. The slow splash appears at the outer pore edge, but tends to travel sideways rather than upward. It broadens towards the inner pore edge (see side view image), and remains thin in the other horizontal dimension (see bottom view image). This splash is the one studied by \citet{Roisman2010}, who restricted themselves to impacts near $1.0$ millimetre-size pores for $\del > 1$, \ie, for no overlap between droplet and pore, and is simply caused by liquid deflected by the pore edge. In other words, we believe that this splash has a similar origin as the one observed for the pit. The disappearance of the splash for large $\del$ ($>1.5$) is likely due to the fact that the spreading liquid looses its kinetic energy to surface energy \citep{Clanet2004}. On the other boundary, $\del<1$, not all downward motion will be transferred to (horizontal) spreading and this motion is not opposed and redirected by the air present in the pore, as in the case of the pit. Therefore, the main flow will become downward, all the way through the pore. 

The other splash regime is found between $0.5< \del < 0.7$, and has sharp transitions towards planar spreading at both boundaries. Here we observe thin splashes, which reach higher velocities than the slow splash (see \textsection \ref{subset:splashvel_pore} and figure \reff{splashvel_pore}). These thin splashes are created when there is already partial overlap between drop and pore, such that quickly after impact the pore edge is partially wetted. As there is no pressure to oppose it, the liquid at the edge can flow into the pore. Therefore, we conjecture that liquid entering from different sides collides inside the pore, leading to a focussed splash at the outer edge.

The intermediate region of $0.7<\del<1.0$, where there is only planar spreading or a small blob visible, suggests that there is a delicate balance between spreading and downward motion. This balance is responsible for the transitions between splashing, spreading and thin splashing. Furthermore, if we compare the thin splash with the splash above a pit for the same $\del$-region, the trend in the velocity, volume, and time of splash formation are all different.  Moving to larger overlap, \ie, $\del < 0.5$, liquid simply enters into the pore, as also observed by \citet{Delbos2010} for full overlap.

\subsubsection{Comparing pit and pore}
The key difference between the pore and the pit is the difference in interaction between liquid and air, which becomes clear when comparing figure \reff{phase_pore} with figures \reff{phen1} and \reff{phen23}. The bottom view images in figure \reff{phase_pore}\textit{d} show that the pore acts as an obstacle for the spreading liquid, which it does not overcome (at least not within the investigated parameter space). There is always a dry area behind the pore, often with opening angles larger than $90^\circ$, measured from the centre of the pore, indicated in figure \reff{phase_pore}\textit{d}. For the pit (figure \reff{phen1} and \reff{phen23}), the spreading liquid not only encloses the pit, but for small and large overlap clearly overcomes the disruption. As soon as the pit is sealed, the air inside the pit will behave as a kind of substrate over which spreading is possible. For the pore, however, air can always move out at the bottom, and hence there is no resistance for liquid to enter. 

Furthermore, even when the pit is not yet sealed by the droplet the air will tend to leave it. As the air can be assumed to be an incompressible fluid, the volume of water that enters the pit will be equal to the volume of air leaving. Therefore, the liquid will already be deflected before the pit is closed off. We can make an estimate of a necessary pit depth, for which the air does not come out immediately and thus, compressibility effects become relevant. The timescale for the pressure wave to bounce back from the pit bottom is given by $t_c = 2 Z_p / c$, with $c$ the velocity of sound in air, and should be comparable to the impact time $\tau = D_d / (2 U_i) = 0.5$~ms. This results in a pit depth of at least $85$~mm before one may expect any pore-like splash behaviour for pits. The difference in air flow explains why the splashes for pit and pore differ in velocity (and volume), even in the case that there is no overlap between the impacting droplet and pit or pore (figure \reff{phen1} and last image in figure \reff{phase_pore}\textit{d}). Furthermore, it provides a plausible reason that pit splashes are always observed whereas for pores, they only occur for a narrow $\del$ range.

\subsection{Splash velocity \label{subset:splashvel_pore}}
\begin{figure*}\begin{center}
	\includegraphics[width=1.0\textwidth]{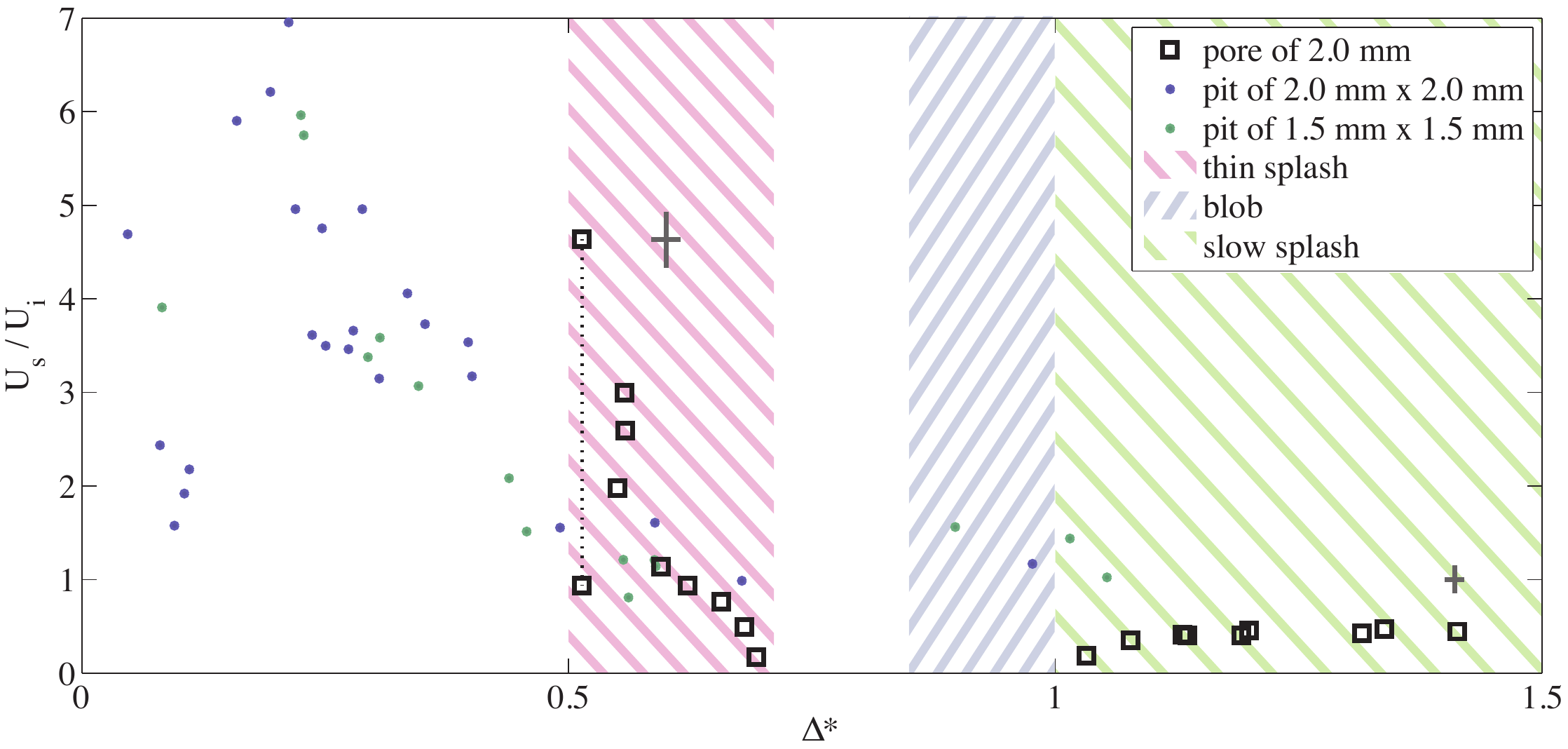}
	\caption{The splash velocity normalised by the impact velocity  $U_s/U_i$ versus the dimensionless distance $\del$ for impacts near a pore ($\Box$). The shaded regions are extracted from figure \reff{phase_pore} to indicate where splashing is expected. The pit data ($\bullet$) is added as a reference. Error bars are added indicating the typical error in the two splashing regions.
	}
	\label{fig:splashvel_pore}
\end{center}\end{figure*}

The splash velocity in the two splash regions is measured and reported in figure \reff{splashvel_pore}. In the (slow) splashing regime, the splash velocity is rather constant around a value of $0.4U_i$. In the thin splashing regime, however, the splash velocity increases with decreasing distance. At $\del \approx 0.5$, two data points are connected with a dashed line. The higher data point corresponds to a fast droplet that shoots out of the pore, of which we could not determine the exact origin, but probably originates from the tip of the splash. A much slower splash appears $0.3$ ms later, represented by the lower data point. The error bars in the figure differ slightly for the two splash regimes. For no overlap, the locations of droplet and pore can be determined with more certainty and the splash fluctuates less, resulting in a smaller error.

The splash velocity for impacts near pores is, in general, smaller than for impacts near pits. The maximum splash velocity ratio ($U_s / U_i$) is at least $1.5$ times smaller, and for the slow splashing regime, \ie, when there is no overlap, it is even $2.5$ times smaller. This demonstrates again, that the air present in the pit plays a very important role and that it influences both the size of the splashing regime and the splash velocity. There is a small region of $0.5<\del<0.55$ where the thin splash is faster than the pit splash, but overall velocities of the pit splashes are much larger. Furthermore, the splash becomes very thin, and thus has little momentum, again pointing to a different origin for the thin splash in comparison to both the pit splash and the slow splash.

\section{Conclusions \label{sec:the_end}}
We have found experimentally that the behaviour resulting from the impact of a water droplet near a closed pit and an open-ended pore greatly differs. Furthermore, we observed distinct impact phenomena when the amount of overlap between the droplet and the hole is varied (table \ref{tab:phen_explained}). 

For impact near a pit these phenomena are, from large to intermediate to small distance $\del$: a splash only, both a splash and a jet, and both a splash and an air bubble respectively (figures \reff{phen1} and \reff{phen23}). Besides that, we found that splashing is most pronounced for pit sizes around $1.5 - 2.0$ mm in diameter and depth, reaching splash velocities up to seven times the impact velocity, see figure \reff{velsplash}. The size of the jetting regime, in figure \reff{phasediagram}, clearly narrows as the pit depth is increased, and completely disappears when the depth is infinite (\ie, the pore case). 

For impact near pores, we only observe splashing. This stands to reason, since the formation of the jet requires the presence of the flow-deflecting pit bottom and for the emergence of an air bubble an enclosed volume of air is needed, both of which are not present in the case of a pore. Surprisingly, we find two distinct splashing regimes, with horizontal spreading in-between (figure \reff{phase_pore}). One regime is for $1.0<\del<1.5$, \ie, when there is no overlap between droplet and pore, and for which a slow, sideways splash is observed. The other regime is at $0.5<\del< 0.7$, where a thin, fast splash is created, which is less pronounced than for the pit case (see figure \reff{splashvel_pore}). 
More research is needed to determine the origin of this splash. We believe the first splash has the same origin as for the pit splash: spreading liquid above the hole is deflected at the outer hole edge.

We state that, beside the presence of a bottom, the role of air is crucial for the observed differences between the impact phenomena near a pit and a pore. As soon as the pit is sealed off by the spreading liquid, the air inside will oppose liquid to enter and therefore, promote spreading and splashing. In order for liquid to enter the pit, air must be displaced by the ensuing pressure build-up inside the pit. Air might then leave either as a distinct air bubble or along with the liquid that jets out of the cavity. A similarity for impact near a pit and a pore is the splash that is observed. However, for the pore, the splash is only visible in a narrow $\del$ range, whereas for the pit, the full investigated $\del$ range is covered. We believe that this difference will at least be observed as long as the air inside the pit can be assumed to be incompressible, which we estimated to be valid for pit depths smaller than $85$~mm. 

Our whole study is focussed on the inertial regime, with $\mbox{\textit{We}} \approx 370$ (and $\mbox{\textit{Re}} \approx 9 \cdot 10^3$) in almost all cases. We believe that the phenomena discussed in this work can be observed for the whole inertial regime (as long as no corona splashing occurs). Furthermore, the splash Weber number can be estimated as $30$, also indicating that the phenomenon can still appear when \textit{e.g.}, surface tension is changed. The jet, however, we expect to be mainly affected by a change in the impact Weber number, as the jet velocity reduces towards the sides of the $\del$ range shown in figure \reff{phasediagram}. Hence, we expect that a reduced $\mbox{\textit{We}}$ number will give a smaller $\del$ range for which jets are observed.

This work is supported by FOM and NWO through a VIDI Grant No. 68047512. 


\end{document}